\documentclass[prd,superscriptaddress,altaffilletter,showpacs,nofootinbib]{revtex4}
\usepackage{amssymb}
\usepackage{amsmath}
\usepackage{graphicx}
\usepackage{subfigure}
\usepackage{epsfig}
\usepackage{epstopdf}
\usepackage[usenames,dvipsnames]{color}
\usepackage{dcolumn}
\usepackage{bm}
\usepackage{color}

\begin{document}

\title{Coulomb's law corrections and fermion field localization in a tachyonic de Sitter thick braneworld }
\author{R. Cartas-Fuentevilla}\email{rcartas@ifuap.buap.mx }
\affiliation{Instituto de F\'{\i}sica, Benem\'erita Universidad Aut\'onoma de Puebla, Apdo. postal J-48, 72570 Puebla, Pue.,
M\'exico}
\author{Alberto Escalante}\email{aescalan@ifuap.buap.mx }
\affiliation{Instituto de F\'{\i}sica, Benem\'erita Universidad Aut\'onoma de Puebla, Apdo. postal J-48, 72570 Puebla, Pue.,
M\'exico}
\author{Gabriel Germ\'an}\email{gabriel@fis.unam.mx}
\affiliation{Instituto de Ciencias F\'{\i}sicas, Universidad Nacional Aut\'onoma de M\'exico,
Apdo. Postal 48-3, 62251 Cuernavaca, Morelos, M\'exico,}
\author{Alfredo Herrera-Aguilar}\email{alfredo.herrera.aguilar@gmail.com}
\affiliation{Instituto de F\'{\i}sica, Benem\'erita Universidad Aut\'onoma de Puebla, Apdo. postal J-48, 72570 Puebla, Pue.,
M\'exico}
\affiliation{Departamento de F\'{\i}sica, Universidad Aut\'onoma Metropolitana Iztapalapa,
San Rafael Atlixco 186, CP 09340, M\'exico D. F., M\'exico.}
\affiliation{Instituto de F\'{\i}sica y Matem\'{a}ticas, Universidad Michoacana de San Nicol\'as de Hidalgo,\\
Edificio C-3, Ciudad Universitaria, CP 58040, Morelia, Michoac\'{a}n, M\'{e}xico.}
\author{Refugio Rigel Mora-Luna}\email{rigel@fis.unam.mx}
\affiliation{Instituto de Ciencias F\'{\i}sicas, Universidad Nacional Aut\'onoma de M\'exico,
Apdo. Postal 48-3, 62251 Cuernavaca, Morelos, M\'exico,}

\date{\today}

\begin{abstract}
In this work, following recent studies which show that it is possible to localize gravity as well as scalar and gauge vector fields 
in a tachyonic de Sitter thick braneworld, we investigate the localization of fermion fields in this model. In order 
to achieve this aim we consider the Yukawa interaction term between the fermions and the tachyonic condensate scalar field 
$MF(T)\Psi\bar\Psi$ in the action and analyze four different cases corresponding to distinct tachyonic functions $F(T(w))$. 
The only condition that this function must satisfy in order to yield 4D chiral fermions upon dimensional reduction is to be odd in $w$.
These functions lead to a different structure of the respective fermionic mass spectrum. 
In particular, localization of the massless left--chiral fermion zero mode is possible for three of these cases. 
%
%
We further analyze the phenomenology of the Yukawa interaction among fermion fields and gauge bosons localized on the brane 
and obtain the crucial and necessary information to compute the corrections to Coulomb's law coming from massive KK vector 
modes in the non--relativistic limit. These corrections are exponentially suppressed due to the presence of the mass gap in the 
mass spectrum of the bulk gauge vector field. From our results we conclude that corrections to Coulomb's law in the thin brane limit 
have the same form (up to a numerical factor) as far as the left--chiral massless fermion field is localized on the brane. Finally we compute 
the corrections to the Coulomb's law for a more general case, on a thick brane scenario, for which we can do estimates consistent with 
brane phenomenology, i.e. we found that the predicted corrections to the Coulomb law in our model, which are well bounded by the 
observed experimental photon mass, is far beyond its upper bound, positively testing the viability of our tachyonic braneworld.
\end{abstract}




\maketitle

\section{Introduction}

\label{sec:intro}

In recent decades, the emergence of phenomenological braneworld models \cite{RubakovPLB1983136}--\cite{1004.3962} suggested the existence of extra compact and extended dimensions. Within the framework of these models it was possible to geometrically reformulate the gauge hierarchy problem \cite{ADD}--\cite{rs}, to address 
the cosmological constant problem (see, \cite{RubakovPLB1983136},\cite{CosmConst}, for instance) and to localize the known matter fields of the standard model in braneworld scenarios with a simple dimensional reduction \cite{BajcPLB2000}--\cite{corradini}, among other issues. Afterwards, due to the intrinsic singularities that braneworld models possess at the position of the branes, there were proposed several scenarios in which the fifth dimension was modeled by bulk scalar fields, extending the idea of thin branes to thick brane configurations \cite{deWolfe}--\cite{HRMGR}. Recently, thick brane models have been proposed in gravity minimally or non-minimally coupled to scalar fields originating from supergravity theories which can be modeled by sigma models, opening thereby the possibility of linking the phenomenology of thick branes to more fundamental theories \cite{Nejat}. This research line pretends to
understand the standard model physics from a higher dimensional point of view in order to address, reformulate and/or solve several open problems such as the gauge hierarchy problem. As a primary requirement of consistency, these models need to localize not only gravity, but also scalar, vector (gauge), and spinor fields on the brane. 

In this work we will focus on studying the localization of spin--$1/2$ fermions on the braneworld generated by a tachyonic condensate scalar field along with 5D gravity (see \cite{NonLocalizedFermion2},\cite{NonLocalizedFermion3},\cite{HRMGR} for braneworld models of this type) using the conventional mechanism that employs a Yukawa coupling between the fermion and the background scalar field, with the interaction term restricted to be an odd function, and will present positive results. Noteworthy recent works reported a new mechanism for localizing fermions with a different Yukawa  interaction between the background scalar fields and the bulk fermion, where the scalar function can be an even function \cite{YXuFWei}--\cite{HQuChu}. The action for the tachyonic scalar field which models the fifth dimension of our work was originally proposed in \cite{sen}--\cite{sen3}. The introduction of this tachyonic field in the thin braneworld paradigm was proposed in \cite{NonLocalizedFermion2}, however, the corresponding 5D spacetime possesses physical singularities at the place where the branes are positioned. A further development of this model was presented in \cite{NonLocalizedFermion3}, where it was shown that it is not possible to localize both gravity and matter fields on the braneworld due to the shape of the used warp factor. A thick braneworld generalization of this model was presented in \cite{HRMGR} and it was shown that 4D gravity can be localized on it. Moreover, it was proved that the relevant field configuration which gives rise to the braneworld model is \emph{stable} under small scalar fluctuations when the gradient of the tachyonic field is small. The scalar curvature which corresponds to this model is positive definite and asymptotically approaches a 5D Minkowski space-time, in contrast with all of the models, to the best of our knowledge, previously reported in the literature. Thus, this model is completely regular and asymptotically flat, instead of (A)dS$_5$. Quite recently it was also reported that in this braneworld it is possible to localize different matter fields as gauge vector fields \cite{AHAgauge} and massive (self--interacting) scalar fields \cite{AHAscalar}. In both of these cases the spectrum for the massive KK modes presents a mass gap which allows us to study in a better way the physics of the massless bound states, specially within the context of computing the higher dimensional corrections to the Coulomb's law that come from the interaction between fermions and gauge bosons localized on the same brane. Thus, the present tachyonic scalar field braneworld turns out to be interesting from the phenomenological viewpoint compared to previous works since it allows us to localize gravity as well as massive scalars, gauge vector fields and fermions.

The paper is organized as follows: Section II contains a brief review of the
tachyonic scalar field braneworld model \cite{HRMGR}; Section III presents four cases where we discuss the problem of fermion
localization on the brane; in Section IV we compute corrections to Coulomb's law coming from the extra dimensional
world for two point fermions interacting with a gauge boson localized on the brane. Finally, we conclude in Section
V with a general discussion of our results.

\section{The thick braneworld model }
The 5D action for the thick braneworld model generated by a tachyon condensate scalar field  reads
\cite{HRMGR}
\begin{equation}
S = \int d^5 x \sqrt{-g} \left(\frac{1}{2\kappa_5^2} R -
\Lambda_5\right) - \int d^{5}x \sqrt{-g}
V(T(w))\sqrt{1+g^{AB}\partial_{A} T(w)\partial_{B} T(w)}, \label{action}
\end{equation}
where $R$ is the 5D scalar curvature, $\Lambda_{5}$ is the bulk
cosmological constant, and $\kappa_5^2=8\pi G_5$ with $G_5$ being the 5D Newton constant. Here we use the signature
$(-++++)$ and the Riemann tensor, defined as follows
$R_{MNPQ}=\frac{\Lambda_5}{6}\left(g_{MP}g_{NQ}-g_{MQ}g_{NP}\right)$,
gives rise to the Ricci one $R_{NQ}=R^M_{NMQ}$ upon contraction of
its first and third indices, where $M,N,P,Q=0,1,2,3,5.$  The corresponding Einstein equations with a cosmological constant in five
dimensions are
\begin{eqnarray}\label{EinsteinEq_5d}
G_{AB} = - \kappa_5^2 ~\Lambda_5 g_{AB} + \kappa_5^2
~T_{AB}^\emph{{bulk}}.
\end{eqnarray}
The 5D metric ansatz compatible with an induced  flat FRW metric on the 3--brane 
has the form
\begin{eqnarray}
\label{metricw}
 ds^2 = e^{2f(w)} \left[- d t^2 + a^2(t) \left(d x^2 + d y^2 + d
z^2 \right)+ d w^2 \right],
\end{eqnarray}
where $\text{e}^{2f(w)}$ and $a(t)$ are the warp factor and the
scale factor of the brane, respectively, and $w$ stands for the extra extended
dimensional coordinate. The matter field equation is obtained by
variation of the 5D action (\ref{action}) with respect to the condensate tachyonic field. It is expressed
as follows:
\begin{equation}
\partial_{M}\left[\frac{\sqrt{-g} V(T) \partial^M T}{\sqrt{1+  (\nabla T)^2}}
 \right ]
- \sqrt{-g} \sqrt{1+  (\nabla T)^2} \frac{\partial V(T)}{\partial T}
= 0.
\end{equation}
The solution for the metric coefficients in (\ref{metricw}), i.e. for
the scale and warp factor, respectively reads
\begin{equation}
a(t)=e^{H\,t}, \qquad  \qquad
f(w)=-\frac{1}{2}\ln\left[\frac{\cosh\left[\,H\,(2w+c)\right]}{s}\right],
\label{scalewarpfactors}
\end{equation}
indicating a de Sitter symmetry induced on the 3--brane; the tachyon condensate scalar field has the form
\begin{eqnarray}
T(w) &=& \pm \ b\
\mbox{arctanh}\left[\frac{\sinh\left[\frac{H\,\left(2w+c\right)}{2}\right]}
{\sqrt{\cosh\left[\,H\,(2w+c)\right]}}\right]  \nonumber \\
&=& \pm \ b~\mbox{arcsinh}\left[\text{tanh}(Hw) \right], \label{Tw}
\end{eqnarray}
while the  tachyon condensate potential is given by
\begin{eqnarray}
&&V(T) = - \Lambda_5\
\mbox{sech}\left(T/b\right)
\sqrt{6\ \mbox{sech}^2\left(T/b\right)-1} = \nonumber \\
&& - \Lambda_5\ \sqrt{\Big(1+\mbox{sech}\left[\,H\,(2w+c)\right]\Big)\left(1+\frac{3}{2}\ \mbox{sech}\left[\,H\,(2w+c)\right]\right)},
\label{VT}
\end{eqnarray}
where $H$ and $c$ are integration constants, and we have set 
\begin{equation}
s=-\frac{6H^2}{\kappa_5^2\,\Lambda_5} 
\qquad 	\quad 
\mbox{and}
\qquad \quad
b=\sqrt{\frac{-3}{2\,\kappa_5^2\,\Lambda_5}},
\label{s}
\end{equation}
with an arbitrary negative bulk cosmological constant $\Lambda_5<0$.

\section{Localization of Spin--1/2 fermion fields}

In this section we shall investigate the localization of spin--1/2 fermion
bulk matter fields on a tachyon condensate de Sitter thick braneworld model by considering a very weak
gravitational interaction between gravity and the fermionic fields, so that the brane solution given in the
previous section remains valid even in the presence of generalized 5D bulk matter. As a generic feature, 
the 5D profile of the fermion fields obey a Schr\"{o}dinger equation when assuming that the corresponding 
4D Dirac equations are satisfied.
The mass spectra of the fermion fields on the de Sitter thick brane will also be discussed by  analyzing the
potential of, and by analytically solving the corresponding Schr\"{o}dinger equation for their KK massive 
modes related to the 4D fermionic fields in four different cases. 

In 5D spacetime fermions are four--component spinors and their Dirac structure can be
described by $\Gamma^M= e^M_{~\bar{M}} \Gamma^{\bar{M}}$ with $e^M_{~\bar{M}}$ being the vielbein
and $\{\Gamma^M,\Gamma^N\}=2g^{MN}$. In this section $\bar{M}, \bar{N}, \cdots =0,1,2,3,5$ and
$\bar{\mu}, \bar{\nu}, \cdots =0,1,2,3$ denote the 5D and 4D local
Lorentz indices, respectively, and $\Gamma^{\bar{M}}$ are the gamma matrices in 5D
flat spacetime. In our set-up, the vielbein is given by
\begin{eqnarray}
e_M ^{~~\bar{M}}= \left(%
\begin{array}{ccc}
  \text{e}^{f} \hat{e}_\mu^{~\bar{\nu}} & 0  \\
  0 & \text{e}^{f}  \\
\end{array}%
\right),\label{vielbein_e}
\end{eqnarray}
$\Gamma^M=\text{e}^{-f}(\hat{e}^{\mu}_{~\bar{\nu}}
\gamma^{\bar{\nu}},\gamma^5)=\text{e}^{-f}(\gamma^{\mu},\gamma^5)$,
where $\gamma^{\mu}=\hat{e}^{\mu}_{~\bar{\nu}}\gamma^{\bar{\nu}}$,
$\gamma^{\bar{\nu}}$ and $\gamma^5$ are the usual flat gamma
matrices in the 4D Dirac representation. The generalized Dirac
action of a spin--1/2 fermion with a mass term can be expressed as
\cite{OdaPLB2000113}
\begin{eqnarray}
S_{\frac{1}{2}} = \int d^5 x \sqrt{-g} \left[\bar{\Psi} \Gamma^M
          \left(\partial_M+\omega_M\right) \Psi
          - M  \bar{\Psi}F(T)\Psi\right]. \label{DiracAction}
\end{eqnarray}
Here $\omega_M$ is the spin connection defined as $\omega_M=
\frac{1}{4} \omega_M^{\bar{M} \bar{N}} \Gamma_{\bar{M}}
\Gamma_{\bar{N}}$ with
\begin{eqnarray}
 \omega_M ^{\bar{M} \bar{N}}
   &=& \frac{1}{2} {e}^{N \bar{M}}\left(\partial_M e_N^{~\bar{N}}
                      - \partial_N e_M^{~\bar{N}}\right)
    - \frac{1}{2} {e}^{N\bar{N}}\left(\partial_M e_N^{~\bar{M}}
                      - \partial_N e_M^{~\bar{M}}\right)  \nonumber \\
   && - \frac{1}{2} {e}^{P \bar{M}} {e}^{Q \bar{N}}\left(\partial_P e_{Q
{\bar{R}}} - \partial_Q e_{P {\bar{R}}}\right) {e}_M^{~\bar{R}},
\end{eqnarray}
and $F(T)$ is an arbitrary general scalar function of the tachyon condensate scalar field. 
We will discuss about the properties of the scalar
function $F(T)$ later in the context of the localization of KK
fermion modes. The non--vanishing components of the spin connection
$\omega_M$ for the background metric (\ref{metricw}) has the form
\begin{eqnarray}
  \omega_\mu =\frac{1}{2}(\partial_{w}f) \gamma_\mu \gamma_5
             +\hat{\omega}_\mu, \label{spinConnection}
\end{eqnarray}
here $\hat{\omega}_\mu=\frac{1}{4} \bar\omega_\mu^{\bar{\mu}
\bar{\nu}} \Gamma_{\bar{\mu}} \Gamma_{\bar{\nu}}$ is the spin
connection derived from the metric
$\hat{g}_{\mu\nu}(x)=\hat{e}_{\mu}^{~\bar{\mu}}(x)
\hat{e}_{\nu}^{~\bar{\nu}}(x)\eta_{\bar{\mu}\bar{\nu}}$. Thus, the
equation of motion corresponding to the variation of the action (\ref{DiracAction}) whit respect to $\bar\Psi$
can be written as
\begin{eqnarray}\label{5Ddiracequation}
 \left[ \gamma^{\mu}(\partial_{\mu}+\hat{\omega}_\mu)
         + \gamma^5 \left(\partial_w  +2 \partial_{w} f \right)
         -\text{e}^f MF(T)
 \right ] \Psi =0, \label{DiracEq1}
\end{eqnarray}
where $\gamma^{\mu}(\partial_{\mu}+\hat{\omega}_\mu)$ is the 4D Dirac
operator on the brane.

Next, we will investigate the KK modes for the 5D Dirac equation
(\ref{DiracEq1}), and write the spinor in terms of 4D
effective fields. On account of the fifth gamma matrix $\gamma^{5}$,
we anticipate the left-- and right--handed projections of the
4D part to behave differently. We shall consider the following ansatz for the general chiral decomposition
in (\ref{DiracEq1}): 
\begin{equation}
\label{varsep}
 \Psi= \text{e}^{-2f}\left(\sum_n\Psi_{Ln}(x) L_n(w)
 +\sum_n\Psi_{Rn}(x) R_n(w)\right),
\end{equation}
where $\Psi_{Ln}(x)=-\gamma^5 \Psi_{Ln}(x)$ and
$\Psi_{Rn}(x)=\gamma^5 \Psi_{Rn}(x)$ are the left-handed and
right-handed components of a 4D Dirac field, respectively. Further, we shall
assume that $\Psi_{Ln}(x)$ and $\Psi_{Rn}(x)$ satisfy the 4D Dirac
equations. Therefore the KK modes $L_{n}(w)$ and $R_{n}(w)$ should
satisfy the following coupled equations:
\begin{subequations}\label{CoupleEq1}
\begin{eqnarray}
 \Big(\partial_w
                  + \text{e}^f MF(T) \Big)L_n(w)
  &=&  ~~m_n R_n(w), \label{CoupleEq1a}  \\
 \Big(\partial_w
                  - \text{e}^f MF(T) \Big)R_n(w)
  &=&  - m_n L_n(w). \label{CoupleEq1b}
\end{eqnarray}
\end{subequations}
where $m_n$ is the fermionic 4D mass arising from the separation of variables (\ref{varsep}). From the above coupled 
equations, we can obtain the Schr\"{o}dinger--like equations for the left-- and right--chiral KK modes of fermions:
\begin{eqnarray}
  \Big(-\partial^2_w + V_L(w) \Big)L_n
            &=&m_{n}^{2} L_n,~~
   \label{SchEqLeftFermion}  \\
  \Big(-\partial^2_w + V_R(w) \Big)R_n
            &=&m_{n}^{2} R_n,
   \label{SchEqRightFermion}
\end{eqnarray}
where the corresponding left and right potentials read
\begin{subequations}\label{Vfermion}
\begin{eqnarray}
  V_L(w)&=& \text{e}^{2f} M^{2}F^{2}(T)
     - \text{e}^{f} f' M F(T) -\text{e}^{f}M\partial_{w}F(T), \label{VL}\\
  V_R(w)&=&   \text{e}^{2f} M^{2}F^{2}(T)
     + \text{e}^{f} f' M F(T) + \text{e}^{f}M\partial_{w}F(T). \label{VR}
\end{eqnarray}
\end{subequations}

 We can perform a dimensional reduction on (\ref{DiracAction}) in order to obtain the standard model 4D action for
a massless fermion and a series of massive chiral fermions
\begin{eqnarray}
 S_{\frac{1}{2}} &=& \int d^5 x \sqrt{-g} ~\bar{\Psi}
     \left[ \Gamma^M (\partial_M+\omega_M)
     -MF(T)\right] \Psi  \nonumber \\
  &=&\sum_{n}\int d^4 x \sqrt{-\hat{g}}
    ~\bar{\Psi}_{n}
      \left[\gamma^{\mu}(\partial_{\mu}+\hat{\omega}_\mu)
        -m_{n}\right]\Psi_{n},~~~
\end{eqnarray}
where the following orthonormalization conditions for $L_{n}$ and $R_{n}$ need to be satisfied in order to perform the dimensional reduction:
\begin{eqnarray}
 \int_{-\infty}^{+\infty} L_m L_ndw
   &=& \delta_{mn}, \label{orthonormalityFermionL} \\
 \int_{-\infty}^{+\infty} R_m R_ndw
   &=& \delta_{mn}, \label{orthonormalityFermionR}\\
 \int_{-\infty}^{+\infty} L_m R_ndw
   &=& 0. \label{orthonormalityFermionLR}
\end{eqnarray}
It is easy to see that if one sets $m_n=0$ in the expressions (\ref{CoupleEq1a}) and (\ref{CoupleEq1b}), then one gets an easy way to 
calculate the zero modes for the left an right--chiral fermions
\begin{subequations}\label{zero}
\begin{eqnarray}
  L_0&\propto & e^{-M \int e^{f} F(T(w)) dw}, \label{zerol}\\
  R_0&\propto &  e^{M \int e^{f} F(T(w)) dw}. \label{zeror}
\end{eqnarray}
\end{subequations}

In the next subsections we will investigate four different profiles of the function $F(T)$ in order to localize the 4D fermions on the thick 3-brane.
To achieve this goal we require the effective potentials $V_L $ and $V_R $ to possess a minimum and to be symmetric with respect to their 
position on the thick brane along the extra dimension. Therefore we will demand the function $ F (T(w))$ to be an odd function in $w$.


\subsection{case I: $F(T)=T/b$}

In this case we shall investigate a simple interaction in the action (\ref{DiracAction})  between the 5D fermionic fields and the tachyon condensate scalar field by taking $F(T)=T/b$, 
where we divide the $T$ field by $b=\sqrt{\frac{-3}{2\,\kappa_5^2\,\Lambda_5}}$ in order to make the function $F(T)$ adimensional and make the parameter $M$ to encode all the relevant units for the interaction term of the 5D action. 
For this field configuration we have the following potentials for $L_n$ and $R_n$ 5D Dirac fermions 
\begin{subequations}
\begin{eqnarray}
 V_L&=& M s H\,\text{sech}(H w)\Biggl[\frac{M}{H}\frac{\text{sech}(H w)\,\text{arcsinh}^2[\text{tanh}(H w)] }{1+\text{tanh}^2(H w)} \nonumber \\
        &+& \frac{1}{\sqrt{s }}\left(\text{arcsinh}[\text{tanh}(H w)] \frac{\text{tanh}(2 H w)}{\sqrt{1+\text{tanh}^2(H w)}} - 1\right)\Bigg]\,, \\
V_R&=& M s H\,\text{sech}(H w)\Biggl[\frac{M}{H}\frac{\text{sech}(H w)\,\text{arcsinh}^2[\text{tanh}(H w)] }{1+\text{tanh}^2(H w)} \nonumber \\
         &-& \frac{1}{\sqrt{s }}\!\left(\text{arcsinh}[\text{tanh}(H w)] 
\frac{\text{tanh}(2 H w)}{\sqrt{1+\text{tanh}^2(H w)}} - 1\right)\Bigg]\,.
\end{eqnarray}
\end{subequations}
Both potentials have the same  asymptotic behavior $V_{R,L}(w\rightarrow \pm\infty)\rightarrow0$, the critical value (maximum and minimum) of the  right and left potentials when $w=0$ are, respectively, $V_{R}(w=0)=\sqrt{s}\, H M^2$ and $V_{L}(w=0)=-\sqrt{s}\, H M^2$.	
Both of the potentials have a very complicated form and it is impossible to find an explicit solution for the Dirac fermion fields when trying to analytically solve the 
Schr\"odinger equations (\ref{SchEqLeftFermion})--(\ref{SchEqRightFermion}). However, these potentials do not allow us to localize the fermion zero modes, the form of the $L_0$ can easily be found numerically as show in Fig. \ref{VL}.
While the potential is of volcano type, which allows, in principle, the existence of bound states, the zero mode is not localized on the 3--brane because it asymptotically tends to a positive definite constant, indicating that the bottom of the volcano potential is not deep enough to localize fermion fields.

\begin{figure}[htb]
\begin{center}
\includegraphics[width=7cm]{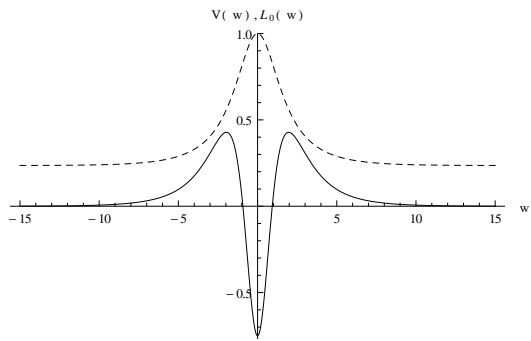}
\end{center}\vskip -5mm
\caption{The profile of the $V_L$ potential (solid black line) and the non--localized left--chiral zero mode $L_0$ (dashed black line) along the
fifth dimension for the case I. Here we have set  $H=1/2$, $M=1$ and $s=1$.} \label{VL}
\end{figure}

The shape of the potential $V_R$ predicts the lack of localized right bound states since it constitutes a barrier potential. 
Fig. \ref{VR} shows the shape of this potential and the massless KK zero mode of the spectrum.

\begin{figure}[htb]
\begin{center}
\includegraphics[width=7cm]{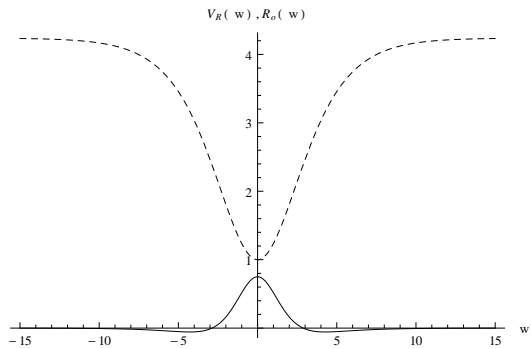}
\end{center}\vskip -5mm
\caption{The profile of the  $V_R$ potential (solid black line) and the non--localized right--chiral massless fermion zero mode $R_O$ (dashed black line) 
along the extra dimension for the case I. Here $H=1/2$, $M=1$ and $s=1$ as well.} \label{VR}
\end{figure}

Thus, for the above analyzed case I there are no, neither left nor right, fermionic bound states localized on the considered 5D braneworld model generated by gravity 
in complicity with the bulk tachyonic scalar field. Therefore, we need to explore more complicated functions $F(T)$ in order to achieve the desired fermion field localization
on the aforementioned braneworld model.

\subsection{case II: $F(T)=\frac{\text{sinh}(2T/b)}{2\left[1-\text{sinh}^{2}(T/b)\right]}
$}

We shall now propose one case in which the left  KK ground state is localized in our braneworld model. 
Therefore, we shall consider a new $ F(T)$ for which we obtain the following expressions for the left and right potentials
\begin{subequations}
\begin{eqnarray}
 V_L &=&MH^2G [MG\,\text{sinh}^2 (H w)-\cosh (H w)], \\
 V_R &=&MH^2G [MG\,\text{sinh}^2 (H w)+\cosh (H w)], 
 \label{potentialsc4}
\end{eqnarray}
\end{subequations}
where $G=\sqrt{\frac{-6}{\kappa^{2}_{5}\Lambda_5}}=\frac{\sqrt{s}}{H}$.

\begin{figure}[htb]
\begin{center}
\includegraphics[width=7cm]{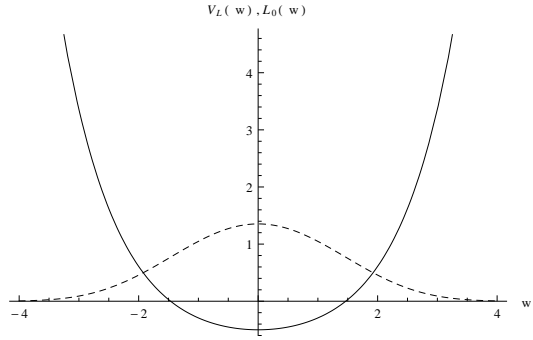}
\end{center}\vskip -5mm
\caption{The profile of the $V_L$ potential (solid black line) and the localized left--chiral zero mode $L_0$ scaled by a factor of 10 (dashed black line) along the
fifth dimension for case II. Here we have set  $H=1$, $M=1/2$ and $s=1/2$.} \label{VL4}
\end{figure}

Both of these potentials have the same asymptotic behaviour $V_{R,L}(w\longrightarrow \pm\infty)\longrightarrow\infty$, giving rise to infinitely high well potentials, which means in turn that the mass spectra of both left-- and right--chiral fermions consists of an infinite set of discrete massive bound states localized on the thick 3--brane.
The critical values of the right and left potentials take place when $w=0$ and are given by $V_{R}(w=0)=H^2 MG$ and $V_{L}(w=0)=-H^2 MG$, respectively. Thus, both of the potentials possess a tower of discrete KK bound states, the only essential difference is that the left--chiral KK fermionic ground state is massless (see Fig. \ref{VL4}), while the right--chiral KK fermionic ground state is a massive one.

The general solution for both the left and right KK bound states can be expressed in terms of confluent Heun functions as follows
\begin{eqnarray}
&&L_{n}=e^{MG\, \cosh (H w)} \left[K_1\, \text{HeunC}\left(4MG,-\frac{1}{2},-\frac{1}{2},2MG,\Omega_{n_-} ,\frac{1}{2}+\frac{1}{2} \cosh (H w)\right)\right.+\nonumber\\
&& \left. K_2\, \sqrt{2+2\cosh (H w)}\, \text{HeunC}\left(4MG,\frac{1}{2},-\frac{1}{2},2MG,\Omega_{n_-} ,\frac{1}{2}+\frac{1}{2} \cosh (H w)\right)\right],
\end{eqnarray}
\begin{eqnarray}
&&R_{n}=e^{MG\, \cosh (H w)} \left[k_1\, \text{HeunC}\left(4MG,-\frac{1}{2},-\frac{1}{2},-2MG,\Omega_{n_+} ,\frac{1}{2}+\frac{1}{2} \cosh (H w)\right)\right.+\nonumber\\
&& \left. k_2\, \sqrt{2+2\cosh (H w)}\, \text{HeunC}\left(4MG,\frac{1}{2},-\frac{1}{2},-2MG,\Omega_{n_+} ,\frac{1}{2}+\frac{1}{2} \cosh (H w)\right)\right],
\end{eqnarray}
where $K_1,K_2,k_1,k_2$ are arbitrary constants and $\Omega_{n_\pm}=\frac{(3 \pm 8 MG) H^2+8 m_n^2}{8H^2}$.

Returning to our goal, we need to know, in particular, the explicit expression for the left and right KK ground states. By proceeding to calculate these expressions as usual we get
\begin{subequations}
\begin{eqnarray}
  L_0&\propto & e^{-MG\,\text{cosh}(H w)}, \label{zerolf2F4}\\
  R_0&\propto & e^{MG\,\text{cosh}(H w)}, \label{zerorf2F4}
\end{eqnarray}
\end{subequations}
implying that just the massless left--chiral fermionic zero mode is localized on the 3--brane.

\begin{figure}[htb]
\begin{center}
\includegraphics[width=7cm]{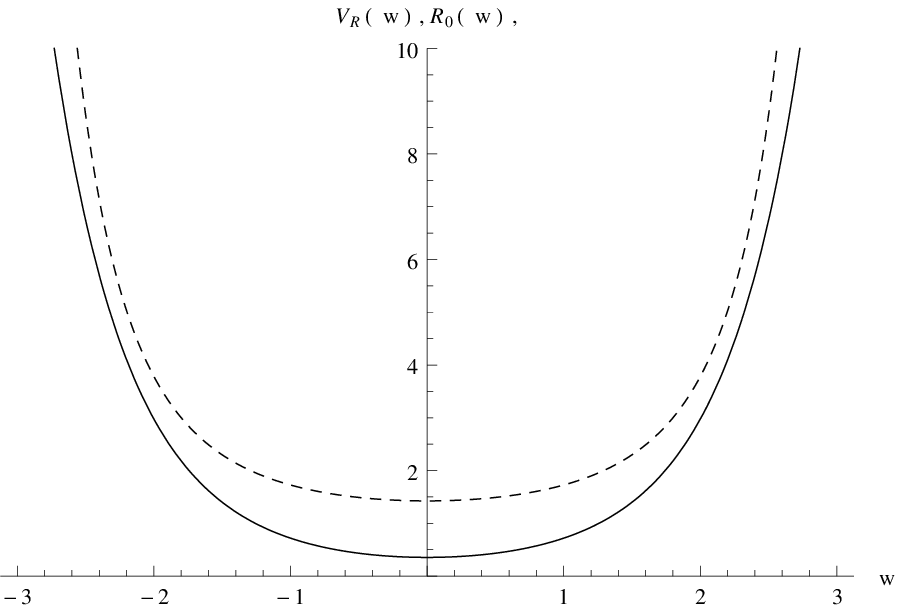}
\end{center}\vskip -5mm
\caption{The profile of the  $V_R$ potential (solid black line) and the non--localized right--chiral zero mode $R_O$ (dashed black line) along the
extra dimension in case II. Here we have set  $H=1.0$, $M=1/2$ and $s=1/2$,} \label{VR4}
\end{figure}

We must emphasize that the shape of the potential $V_R$ predicts the existence of an infinite tower of discrete massive bound states localized on the brane along with the presence of a non--localized massless zero mode.  Fig. \ref{VR4} shows the shape of the right potential and the delocalized massless zero mode from the brane. 

\subsection{case III: $F(T)=\frac{\text{arctanh}\left[\text{sinh}\left(T/b \right)\right]}{\sqrt{2\, \text{sech}^2\left(T/b \right)-1}\ \left(1+\text{arctanh}^2\left[\text{sinh}\left(T/b\right)\right]\right)}
$}

\begin{figure}[htb]
\begin{center}
\includegraphics[width=7cm]{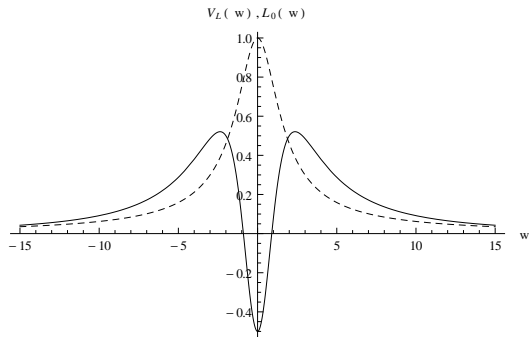}
\end{center}\vskip -5mm
\caption{The profile of the $V_L$ potential (solid black line) and the localized left--chiral ground state $L_0$ (dashed black line) along the
fifth dimension for case III. Here $H=1/2$, $M=1$ and $s=2$.} \label{VLC2}
\end{figure}

\begin{figure}[htb]
\begin{center}
\includegraphics[width=7cm]{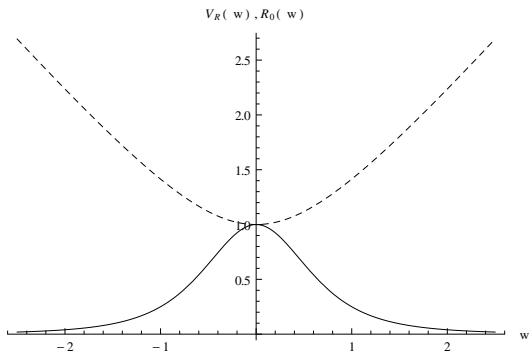}
\end{center}\vskip -5mm
\caption{The profile of the $V_R$ potential (solid black line) and the non--localized right--chiral zero mode $R_0$ (dashed black line) along the
extra dimension for case III. Here we have set  $H=1$, $M=1/2$ and $s=1$.} \label{VRC2}
\end{figure}

In general, the localization of spin--$\frac{1}{2}$ fermions is obtained in a more artisanal way when compared to the localization of gravity, scalar and/or gauge vector fields. This is why we shall undertake the task of finding field configurations with a little whimsical $F(T)$, like the one considered here in case III, that allows us to localize fermion fields on the 3--brane.
For the configuration corresponding to case III we get again a left potential $V_L$ of volcano type (see Fig. \ref{VLC2}), while the shape for the right potential $V_R$ is conceived as a barrier potential as shown in Fig. \ref{VRC2}. The expression for both the left and right potentials reads
\begin{subequations}\label{Vfermion}
\begin{eqnarray}
  V_L(w)&=&  \frac{H^2MG\left[(MG+1)H^2w^2-1\right]}{\left(1+H^2 w^2\right)^2}, \label{VLf}\\
  V_R(w)&=& \frac{H^2MG\left[(MG-1)H^2w^2+1\right]}{\left(1+H^2 w^2\right)^2}. \label{VRf}
\end{eqnarray}
\end{subequations}
where the constant $G$ is defined as in the case II. This two potentials have the same vanishing asymptotic behavior $V_{R,L}(w\longrightarrow \pm\infty)=0$, indicating the lack of a mass gap in their corresponding mass spectra;
the critical values (maximum and minimum) of the right and left potentials is achieved when $w=0$ and are respectively given by $V_{R}(w= 0)=H^2MG $ and 
$V_{L}(w=0)=-H^2MG$.	
Only the left  potential supports a left--chiral zero mode $L_0$ localized on the brane. The volcano potential $V_L$ supports a tower of continuous KK massive modes non--localized 
on the 3-brane. On the other hand, the right potential  $V_R $ represents a barrier potential, a fact which indicates that right--chiral fermions cannot be localized on the 3--brane.

By making use of the relations given by (\ref{zero}) we can easily find expressions for the massless zero modes $R_0$ and $L_0$ supported by the potentials (\ref{Vfermion})
\begin{subequations}
\begin{eqnarray}
  L_0&\propto & e^{-M \int\frac{H w \sqrt{s }}{\left(1+H^2 w^2\right) } dw}=\left(1+H^2 w^2\right)^{-\frac{MG}{2}}, \label{zerolf2}\\
  R_0&\propto &  e^{M \int\frac{H w \sqrt{s }}{\left(1+H^2 w^2\right) } dw}=\left(1+H^2 w^2\right)^{\frac{MG}{2}}. \label{zerorf2}
\end{eqnarray}
\end{subequations}

Moreover, by solving the Schr\"odinger equation corresponding to the left potential $V_L$ (\ref{VLf}) we can also obtain the general solution for the KK excitations with
arbitrary mass and see that  the continuous spectrum of KK massive modes can be expressed in terms of confluent Heun functions in the following form
\begin{eqnarray}
L_{n}&=&C_1 \left(1+H^2 w^2\right)^{1+\frac{MG}{2}} \text{HeunC}\left(0,-\frac{1}{2} ,1+MG ,-\frac{1}{4} \frac{m_{n}^2}{H^2} ,\eta_n ,-H^2 w^2\right)+\nonumber\\
&&C_2 \left(1+H^2 w^2\right)^{1+\frac{MG}{2}} w\, \text{HeunC}\left(0,\frac{1}{2} ,1+MG ,-\frac{1}{4} \frac{m_{n}^2}{H^2} ,\eta_n ,-H^2 w^2\right),
\end{eqnarray}
where $C_1$ and $C_2$ are arbitrary constants, and the parameters $\eta_n$ is given by $\eta_n =\frac{(2+MG)H^2+m_{n}^2}{4H^2}$.

We can conclude that case III yields a left--chiral massless fermion zero mode localized on the 3--brane of our model, along with a continuum 
of massive KK fermionic excitations delocalized from the brane, whereas all right--chiral KK fermionic modes are non--localized on the brane. 


\subsection{case IV: $F(T)=\frac{2\,\text{tanh}(T/b)}{\sqrt{1-\text{sinh}^{2}(T/b)}}
$}

In this case we have for both the left and right potentials a modified P\"oschl--Teller configuration which has been carefully studied in several 
modern physics scenarios. This function $F(T)$ allows us to have KK discrete  and continuous mass spectra separated by a mass gap from the 
massless zero mode. The size of these mass gaps largely depend on the value of 4D and 5D parameters as shown by the following expressions:
\begin{subequations}\label{Vfermion2}
\begin{eqnarray}
  V_L(w)&=& M \left[M s \tanh^2 (2H w)-2H \sqrt{s}\, \text{sech}^2(2H w)\right], \label{VLf2}\\
  V_R(w)&=& M \left[M s \tanh^2 (2H w)+2H \sqrt{s}\, \text{sech}^2(2H w)\right]. \label{VRf2}
\end{eqnarray}
\end{subequations}
By substituting the value of $s$ in (\ref{Vfermion2}) and recalling that $b=\sqrt{\frac{-3}{2\kappa^{2}_{5}\Lambda_{5}}}$ according to (\ref{s}), we can 
recast the potentials $V_{R,L}$ as 
\begin{subequations}\label{Vfermion2D}
\begin{eqnarray}
  V_L(w)&=& 4 M H^2 b\left[ M b \tanh^2 (2H w)- ~\text{sech}^2(2H w)\right], \label{VLf2D}\\
  V_R(w)&=& 4 M H^2b \left[ M b \tanh^2 (2H w)+ ~\text{sech}^2(2H w)\right]. \label{VRf2D}
\end{eqnarray}
\end{subequations}

The asymptotic behaviour for the potentials has the form $V_{R,L}(w\longrightarrow \pm\infty)= M^2\,s=4 M^2 H^2 b^2$ and is positive definite, a fact which in general ensures the 
existence of a mass gap between the bound states of the corresponding mass spectra. The critical values (maximum and minimum) of the right and left potentials when $w=0$ are respectively  $V_{R}(w=0)=4 M H^2 b$ and $V_{L}(w=0)=- 4 M H^2 b$.	
The massless zero modes for both potentials can be written as follows
\begin{subequations}
\begin{eqnarray}
  L_0&\propto & \text{sech}^{Mb}(2H w), \label{zerolf2D}\\
  R_0&\propto &  \cosh^{Mb}(2H w). \label{zerorf2D}
\end{eqnarray}
\end{subequations}
\begin{figure}[htb]
\begin{center}
\includegraphics[width=7cm]{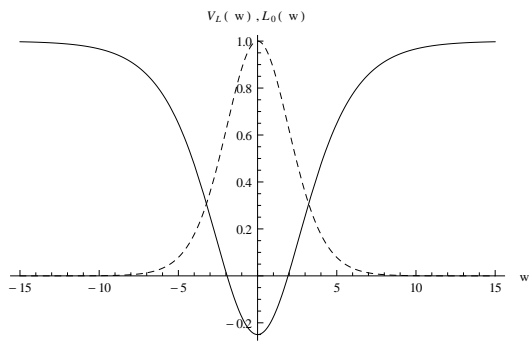}
\end{center}\vskip -5mm
\caption{The profile of the $V_L$ potential
(solid black line) and the localized left--chiral ground state $L_0$ (dashed black line) along the
fifth dimension for case IV. Here we have set  $H=1/4$, $M=1$ and $s=1$.} \label{VLC3}
\end{figure}
\begin{figure}[htb]
\begin{center}
\includegraphics[width=7cm]{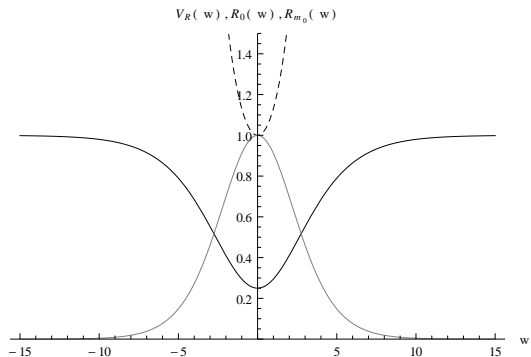}
\end{center}\vskip -5mm
\caption{The profile of the $V_R$ potential (solid black line), the localized right--chiral massive ground state $R_{m_0}$ (gray line), and the non--localized massless zero mode $R_0$ (dashed black line) along the extra dimension for case IV. Here we have also set  $H=1/4$, $M=1$ and $s=1$.} \label{VRC3}
\end{figure}
From these expressions it is clear that just the left--chiral fermion field possesses a localized zero mode on the 3--brane.
The general solution for the $L_{n}$'s is given in terms of hypergeometric functions $_{2}\text{F}_{1}$ and reads
\begin{eqnarray}
\label{LnCaseIIIEven}
 L_{n}&\propto& \cosh^{1+M b}(2Hw)~_{2}\text{F}_{1}
  \left(s_{n},r_{n};\frac{1}{2};-\sinh^{2}(2Hw)\right),
\end{eqnarray}
for even $n$ and
\begin{eqnarray}
\label{LnCaseIIIOdd}
 L_{n}&\propto& \cosh^{1+M b}(2Hw)\sinh(2Hw)~_{2}\text{F}_{1}
  \left(s_{n}+\frac{1}{2},r_{n}+\frac{1}{2};\frac{3}{2};-\sinh^{2}(2Hw)\right),
\end{eqnarray}
for odd $n$, where the parameters $s_{n}$ and $r_{n}$ are given by
\begin{eqnarray}
 s_{n}=\frac{1}{2}(n+1), ~~~~~~~~
 r_{n}=M b-\frac{1}{2}(n-1).
\end{eqnarray}
The number of bound states for the left--chiral fermion $L_n$ is finite, they are labeled by $n=0,1,2,...,< M b$ and 
the corresponding KK mass spectrum is described by 
\begin{eqnarray}
m^{2}_{L_{n}}=4H^2\left(M b-n \right)n.
\label{mnIV}
\end{eqnarray}
If we take into account that, by definition, $b>0$, we can infer that when $ M b< 1 $ there is a single left--chiral bound state, the massless zero $L_0$, as depicted in Fig. \ref{VLC3}, and therefore it is only possible to localize left--chiral fermions on the 3--brane. On the contrary, in order to obtain a finite number of massive KK excited modes we must impose the condition $ M b > 1 $.

For the potential of right-chiral fermions, as shown in (\ref{VRf2D}), it is not possible to localize the massless zero mode. Thus, the only way to ensure the existence of a finite number of
localized bound states for the right--chiral massive fermions consists in imposing the condition $ M b > 1 $.

The general expression for the KK right--chiral bound states in this case is given by \begin{eqnarray}
\label{RnCaseIIIEven}
 R_{n}\propto \cosh^{M b}(2Hw)~_{2}\text{F}_{1}
      \left(\frac{1+n}{2},M b-\frac{1+n}{2};\frac{1}{2};-\sinh^{2}(2Hw)\right),
\end{eqnarray}
for even $n$ and
\begin{eqnarray}
\label{RnCaseIIIOdd}
 R_{n}\propto \cosh^{M b}(2Hw)\sinh(Hz) ~_{2}\text{F}_{1}
      \left(1+\frac{n}{2},M b-\frac{n}{2};\frac{3}{2};-\sinh^{2}(2Hw)\right),	
\end{eqnarray}
for odd $n$.

It is worth emphasizing that the massless zero mode $R_0$ given in (\ref{zerorf2D}) is not a localized fermionic bound state, therefore the ground state for right--chiral fermions corresponds to the first massive bound state (with $n=0$), as illustrated in Fig. \ref{VRC3}, and is denoted by
\begin{eqnarray}
\label{BSm0}
  R_{m_0}\propto \text{sech}^{M b-1}(2Hw),
\end{eqnarray}
where the mass of the first right--chiral KK bound state obeys the following inequality $m^{2}_{R_0}=4H^{2}\left(2M b-1\right)>0$.

The number of bound states for the right--chiral fermion fields inferred from the canonical form of the $V_R$ potential is $n=0,1,2,...,<M b-1$.
For this set of eigenvalues we have the following mass spectrum for the right--chiral fermions $m^{2}_{R_n}=4H^2\left[2M b-\left(n+1\right) \right]\left(n+1\right)$.
We should finally mention that both of the potentials $V_R$ and $V_L$ have a continuous mass spectrum that is achieved when $m_{Ln,Rn}>4M^2H^2b^{2}=M^2\,s$, as it is evident from the asymptotic behaviour of these potentials.

\begin{figure*}[htb]
\begin{center}
\subfigure[$n=1$]{\label{fig_FermionIII_L1}
\includegraphics[width=5cm]{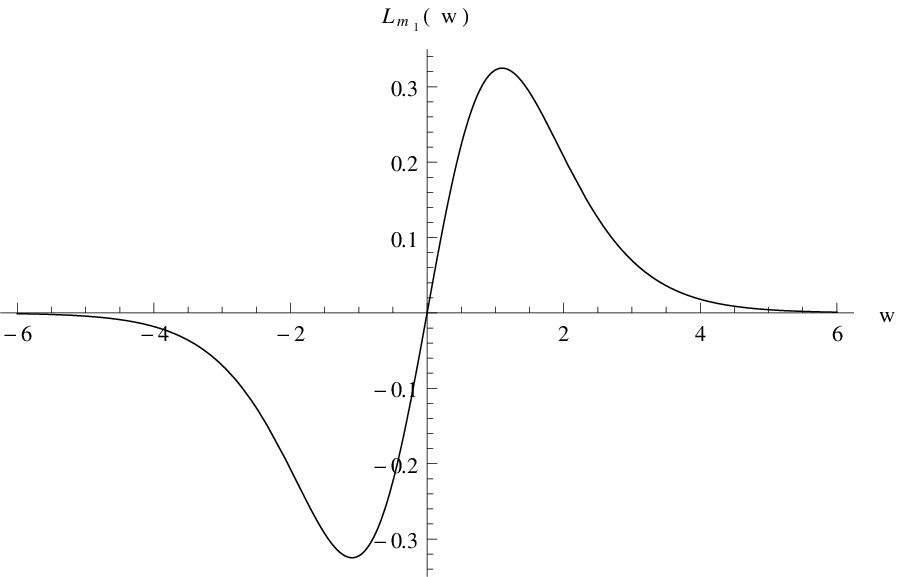}}
\subfigure[$n=2$]{\label{fig_FermionIII_L2}
\includegraphics[width=5cm]{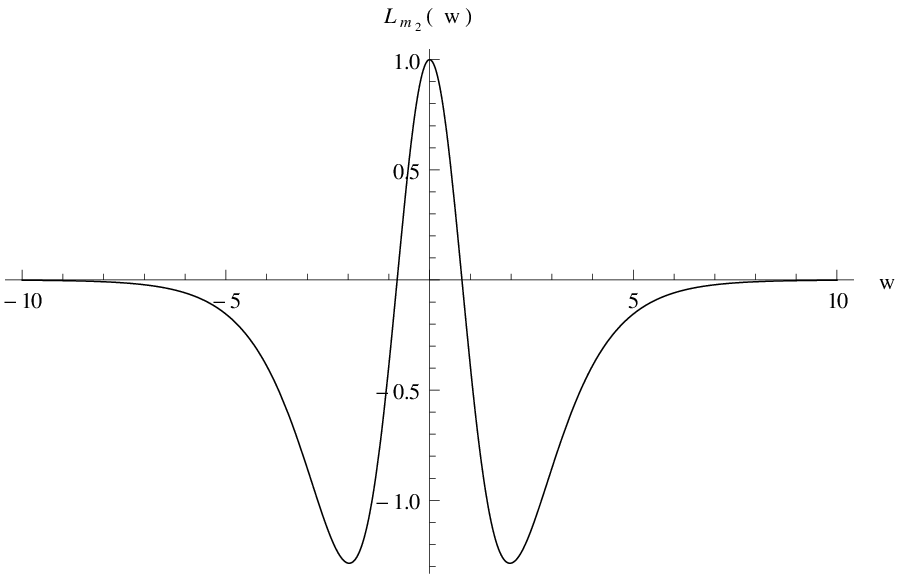}}
\subfigure[$n=1$]{\label{fig_FermionIII_R2}
\includegraphics[width=5cm]{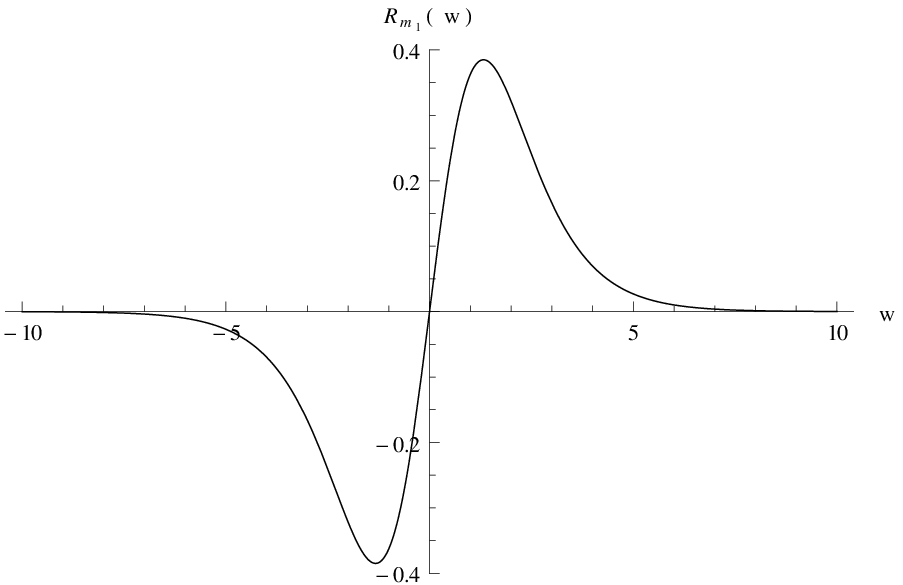}}
\subfigure[$n=2$]{\label{fig_FermionIII_R3}
\includegraphics[width=5cm]{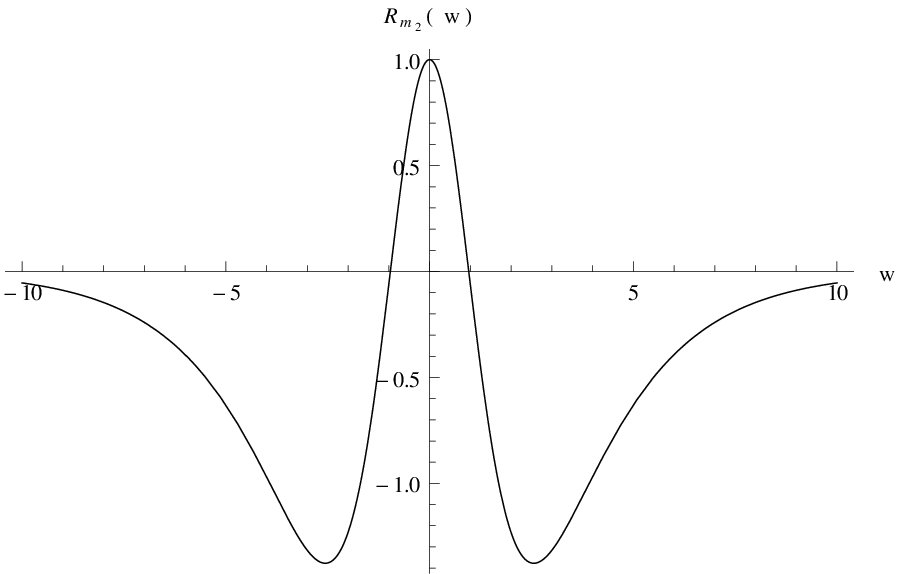}}
\end{center}\vskip -5mm
\caption{The shape of the left-- and right--chiral KK massive modes respectively for $n=1,2$ in case IV. The parameters are set to $M=2$, $b=4$
and $H=1/4$.}
 \label{fig_FermionIII_LnRn}
\end{figure*}

In Fig. \ref{fig_FermionIII_LnRn} we present the profile of left and right--chiral KK massive modes respectively for $n=1,2$ in the above studied case IV.

\section{Corrections to Coulomb's law}
\label{CCL}

In this section we shall compute the Coulomb's law modifications that come from the contribution of the KK massive modes of the bulk gauge vector field. In order to achieve this aim, we shall consider a Yukawa interaction between 5D fermions and gauge bosons which constitutes a generalization of the 4D quantum interacting potential given by 
$L_I= -e \bar{\psi}(x) \gamma^{\mu} A_{\mu}(x) \psi(x)$ with the vertex factor $-i e \gamma^{\mu}$ \cite{peskin}.

Hence, the generalized 5D interaction between fermions and gauge boson reads
\cite{HuangPRD2004}
\begin{eqnarray}
     S_I=  -e_5{\int} d^4 x dw~ \sqrt{-g}~ \bar{\Psi}(x,w)  \Gamma^M A_M(x,w)  \Psi(x,w),
\end{eqnarray}
where $e_5$ is a 5D coupling constant and $A_M(x,w) $ represents the generalized 5D gauge vector field that mediates the interaction between the fermion fields under the gauge condition $A_5=0$ and the KK vector modes decomposition 
\begin{equation}
A_\mu(x,w) = \sum_n a^{(n)}_\mu(x)\rho_n(w)e^{-f/2},
\label{KKdecompAmu}
\end{equation}
where $\rho_n(w)$ is the profile of the massive gauge boson along the fifth dimension.
We shall suppose as well that the 4D fermions are associated to the left--chiral KK massless zero mode $L_0$ of the last three cases considered in the previous section.
The zero mode of this gauge field has recently been shown to be localized on our braneworld model given by (\ref{action})--(\ref{s}) in \cite{AHAgauge}.
Then, by performing the dimensional reduction we can confirm the similarity between the Newton potential for two point particles
interacting with  massive KK tensor modes and the Coulomb potential for two point charges 
interacting with massive KK gauge vector field modes. Let us tart by considering the following action:
\begin{eqnarray}
  S_I&\supset&  \sum \!\!\!\!\!\!\!\! {\int_n} {\int} d^4 x dw~\sqrt{-\hat{g}}~\text{e}^{5f}
      (-e_5)\text{e}^{-2f} \bar{\psi}_0(x) L_0(w)
       \text{e}^{-f}\gamma^{\mu} a_{\mu}^{(n)}(x)
       \text{e}^{-f/2} \rho_n(w)
       \text{e}^{-2f}\psi_0(x) L_0(w) \nonumber \\
  &=&  -e_5 \sum \!\!\!\!\!\!\!\! {\int_n} {\int} dw~\text{e}^{-f/2}~\rho_n(w)  L_0^2(w)
       {\int} d^4 x \sqrt{-\hat{g}}~
       \bar{\psi}_0(x)  \gamma^{\mu} a_{\mu}^{(n)}(x) \psi_0(x) \nonumber \\
  &=& {\int} d^4 x~ \sqrt{-\hat{g}} ~
     \Big\{ -e\bar{\psi}_0(x) \gamma^{\mu} a_{\mu}^{(0)}(x)  \psi_0(x)
            -\sum \!\!\!\!\!\!\!\! {\int_{n\ne 0}}~~e_n \bar{\psi}_0(x) \gamma^{\mu} a_{\mu}^{(n)}(x) \psi_0(x)
     \Big\},
\end{eqnarray}
where the $\sum
\!\!\!\!\!\!{\int}\,_n$ stands for summation or integration (or both) with respect to $n$, depending on the respective discrete or continuos (or mixed) character of the $a_{\mu}^{(n)}(x)$ and $e_n(w)$. By taking into account the form of the gauge vector modes $\rho_{0}(w)$ and $\rho_{n}(w)$ from \cite{AHAgauge} 
\begin{eqnarray}
\rho_{0}(w)&=&\frac{\sqrt{H}(\pi/2)^{1/4}}{2\Gamma(5/4)}\text{sech}^{1/4}(2Hw),\\
\rho_{n}(w)&=&\left[\sum_{\pm} C_{\pm}(\sigma)
 P_{1/4}^{\pm i\sigma}\left(\text{tanh}(2Hw)\right)\right],\label{rhon}
\end{eqnarray}
where $P_{1/4}^{\pm i\sigma}$ are associated Legendre functions of first kind of degree $1/4$ and order $\pm i\,\sigma$
with $\sigma=\sqrt{\frac{m^2}{4H^2}-\frac{1}{16}}$, which imposes the condition $m\ge H/2$, we find the next relations between the couplings $e$, $e_{5}$ and $e_n(w): $
\begin{eqnarray}
e&=&e_5 {\int} dw~\text{e}^{-f/2} ~\rho_0(w)  L_0^2(w)
   =e_5\frac{ \left(2\pi\right)^{1/4}}{\Gamma\left(\frac{1}{4}\right)b^{1/2}}
        {\int} dw~ L_0^{2}(w) = e_5\frac{ \left(2\pi\right)^{1/4}}{\Gamma\left(\frac{1}{4}\right)b^{1/2}}, 
   \label{chare4d}
\end{eqnarray}
where $e$ is the usual 4D charge of the fermion localized on the brane and
$e_n$'s are 4D effective couplings defined as follows
\begin{eqnarray}
e_n &\equiv &  e_5 {\int} dw~\text{e}^{-f/2} ~\rho_n(w)  L_0^2(w) = e \frac{\Gamma\left(\frac{1}{4}\right)b^{1/2}}{\left(2\pi\right)^{1/4}}{\int} dw~\text{e}^{-f/2} ~\rho_n(w)  L_0^2(w).
\label{epsilon_n}
\end{eqnarray}

In the non--relativistic limit the Coulomb potential (and its corrections) between two charged fermions is determined by the KK
photon exchange process and turns out to be
\begin{eqnarray}\label{Potcomplet}	
 V(r) &=& \frac{e^2}{4{\pi}r}+ \int_{m_0}^{\infty}
 dm \frac{e_n^2}{4{\pi}r}\text{e}^{-m r}\noindent\nonumber\\
      &=& \frac{1}{4{\pi}r}
         \left[ e^2 + e^{2}_5 \int_{m_0}^{\infty} dm\ \text{e}^{-m r}
                   \left({\int} dw~\text{e}^{-f(w)/2} ~\rho_n(w)  L_0^2(w)\right)^2
         \right],\label{potentialCoulomb}
\end{eqnarray}
where $m_0=H/2$ is the first excited KK massive mode of the gauge vector field. In this way it is easy to see that the Coulomb
potential arises from the vector zero  mode, while its corrections come from the massive KK vector excitations.

If we pay attention to the formula (\ref{chare4d}) we realize that the existing relationship between the 4D charge $ e $ and the  coupling constant $e_5$  does not depend on the form of the left--chiral KK ground state $L_0$ since it is normalized to unity. On the other hand, the integral in the rhs of the expression (\ref{potentialCoulomb}) for the extra dimensional corrections to the Coulomb potential will always render a constant (which depends on the mass $m$ of the KK gauge field) as far as we suitably define a Dirac delta function with the aid of the left--chiral zero mode $L_0$ in the thin brane limit (see further subsections for concrete examples). Thus, under this definition of the delta function, the squared integral with respect to $w$ in (\ref{potentialCoulomb}), with the prefactor $\text{e}^{-f(w)/2} ~\rho_n(w)$ multiplying a Dirac delta function, will lead to the value $\rho_n(0)^2$ since the Dirac delta function is located at the origin of the fifth dimension $w=0$.
This circumstance makes us conclude, despite the heuristic proposals employed for the function $F(T)$, that the corrections to Coulomb's law associated with the massive KK gauge vector modes in the thin brane limit do not depend on the explicit form of the function $F(T)$ and are, in this sense, model independent as it will be shown in the following examples.

In the following subsections we will analytically compute the Coulomb potential $V(r)$ in the thin brane limit, which is not an easy task, but is still affordable for the three previously studied cases in which the left--chiral massless fermion localization on the 3--brane is feasible.

\subsection{Corrections to Coulomb's law in case II}

In order to compute the Coulomb's law corrections for the case II, let us begin by calculating the 4D effective coupling constants  $e_n$. To do that we shall make use of the fermionic localization mechanism described above with the odd function $F(T)=\frac{\text{sinh}(2T/b)}{1-\text{sinh}^{2}(T/b)}.$ In this case the normalized fermion zero mode reads
\begin{equation}
L_{0}(w) =\left(\frac{H}{\rm{K_{0}}\left(2MG\right)}\right)^{\frac{1}{2}}e^{-MG\,\text{cosh}(H w)},
\end{equation}
where K$_{0}$ is the modified Bessel function of second kind. By substituting the warp factor (\ref{scalewarpfactors}) and the
expression for $\rho_n(w)$ in (\ref{epsilon_n}) we obtain
\begin{eqnarray}
e_n &=& e\frac{\sqrt{H}\, \Gamma\left(\frac{1}{4}\right)}{\sqrt{2}\,(2 \pi )^{1/4}\,\rm{K_{0}}\left(2MG\right)}
{\int} dw~\text{cosh}^{\frac{1}{4}}(2Hw)\,e^{-2MG\, \text{cosh}(H w)}\times\nonumber\\
& & \left[\sum_{\pm} C_{\pm}(\sigma)
 P_{1/4}^{\pm i\sigma}\left(\text{tanh}(2Hw)\right)\right].
\end{eqnarray}
By making use of the following definition of the Dirac delta function\footnote{It is straightforward to check that this definition possesses all the properties of the normalized to unity delta distribution function.} which corresponds to the thin brane limit when $H\to\infty$:
\begin{eqnarray}
\delta (w) = \lim_{H\to\infty} ~\frac{H e^{-2MG\,\text{cosh}(H w)}}{\rm{K_{0}}\left(2MG\right)}, \label{delta4}
\end{eqnarray}
we finally get the following expression for the $ e_n$'s
\begin{eqnarray}
e_n &=& e \frac{ \Gamma\left(\frac{1}{4}\right)}{\sqrt{2H}\,(2 \pi )^{1/4}\,}
\left[\sum_{\pm} C_{\pm}(\sigma)
 P_{1/4}^{\pm i\sigma}\left(0\right)\right].
\end{eqnarray}

Once we have these 4D effective couplings at hand we can write the Coulomb potential as follows
\begin{eqnarray}
V(r) &=& \frac{e^2}{4{\pi}r}\left[1 + \frac{\left[ \Gamma\left(\frac{1}{4}\right)\right]^2}{2H\,\sqrt{2\pi}} \int_{m_0}^{\infty}
dm\ \text{e}^{-m r} \left|\sum_{\pm} C_{\pm}(\sigma) P_{1/4}^{\pm
i\sigma}\left(0\right)\right|^2\right]
\noindent\nonumber\\
      &=& \frac{e^2}{4{\pi}r}\left[1 + \frac{\left[\Gamma\left(\frac{1}{4}\right)\right]^2}{\sqrt{2\pi}\,H}  \int_{m_0}^{\infty}
dm\ \text{e}^{-m r}
\left|\frac{\Gamma\left(1+i\sigma\right)}{\Gamma\left(\frac{3}{8}-
\frac{i\sigma}{2}\right)\Gamma\left(\frac{9}{8}-\frac{i\sigma}{2}\right)}\right|^2\right],
\noindent\label{CCl_Gammas}
\end{eqnarray}
where we have taken into account the fact that the normalization constants for the associated Legendre functions are given by
$C_{\pm}(\sigma)=\frac{\left|\Gamma(1\mp i\sigma)\right|}{\sqrt{2\pi}}$, as well as the following relation
\begin{eqnarray}
P_{\nu}^{\mu}(0)=
\frac{2^{\mu}\sqrt{\pi}}{\Gamma\left(\frac{1-\nu-\mu}{2}\right)\Gamma\left(1+\frac{\nu-\mu}{2}\right)}.
\label{Legendre_0}
\end{eqnarray}
Thus, the Coulomb potential can be written in the form
\begin{eqnarray}
 V(r) = \frac{e^2}{4{\pi}r}\left[1 + \Delta V\right],
\label{VCCl2}
\end{eqnarray}
where the correction $\Delta V$ reads
\begin{eqnarray}
 \Delta V = \frac{\left[\Gamma\left(\frac{1}{4}\right)\right]^2}{\sqrt{2\pi}\,\Gamma \left(\frac{3}{8}\right)^2 \Gamma
\left(\frac{9}{8}\right)^2} 
\frac{e^{-Hr/2}}{Hr}\left(1+{\cal
O}\left(\frac{1}{Hr}\right)\right). \label{CCptl2}
\end{eqnarray}

When performing this computation, in (\ref{CCl_Gammas}) we have expanded the prefactor that multiplies the exponential
function in the integrand with respect to $m_0=H/2$ (which
corresponds to $\sigma=0$) since the corrections to the Coulomb
potential are dominated by the sector of small massive KK vector
modes \cite{PRD0709.3552}.

\subsection{Corrections to Coulomb's law in case III}

We now will calculate the explicit form of the Coulomb potential $V(r)$ following the same procedure as in case II.
Here $F(T)=\frac{\text{arctanh}\left[\text{sinh}\left(T/b \right)\right]}{\sqrt{2\, \text{sech}^2\left(T/b \right)-1}\,
\left[1+\text{arctanh}^2\left(\text{sinh}\left(T/b\right)\right)\right]}$ and for this function the normalizable left--chiral KK massless zero mode is
\begin{equation}
L_{0}(w) =\left[\frac{H~\Gamma\left(MG\right)}{\sqrt{\pi}\,\Gamma\left(MG-\frac{1}{2}\right)}\right]^{\frac{1}{2}}\left(1+w^2H^2\right)^{-\frac{MG}{2}},\qquad MG>\frac{1}{2},
\end{equation}
where the inequality condition follows in order to render a convergent integral.
 
By taking into account the expressions for the warp factor (\ref{scalewarpfactors}) and the gauge function $\rho_n(w)$ (\ref{rhon}) we 
can compute the expression for the coupling constants $e_n$'s (\ref{epsilon_n}) and get the same result as in the previous case:
\begin{eqnarray}
e_n &=& e \frac{ \Gamma\left(\frac{1}{4}\right)}{\sqrt{2H}\,(2 \pi )^{1/4}\,}
\left[\sum_{\pm} C_{\pm}(\sigma) P_{1/4}^{\pm i\sigma}\left(0\right)\right],
\label{e_nIII}
\end{eqnarray}
when defining the Dirac delta function as shown below, in the thin brane limit, when $H\to\infty$:
\begin{eqnarray}
\delta (w) = \lim_{H\to\infty} ~\frac{H~\Gamma\left(MG\right)}{\sqrt{\pi}\,\Gamma\left(MG-\frac{1}{2}\right)}
\left(1+w^2H^2\right)^{-MG}.
\label{delta}
\end{eqnarray}
By substituting the expression (\ref{e_nIII}) into equation (\ref{potentialCoulomb}) we get the same form for the Coulomb potential (\ref{VCCl2}), 
where its correction is again defined as in (\ref{CCptl2}), obtaining the same result as in the above studied case II.

\subsection{Corrections to Coulomb's law for case IV}

We shall further proceed to perform the analytical calculation of $V(r)$ for case IV. Let us compute first the 4D effective couplings $e_n$. In order to achieve this goal we shall make 
use of the function $F(T) = \frac{2\,\text{tanh}(T/b)}{\sqrt{1-\text{sinh}^2(T/b)}}$ in the fermionic localization mechanism for which the normalizable left--chiral zero mode reads
\begin{equation}
L_{0}(w) =\left[\frac{2H\Gamma\left(\frac{1}{2}+M b\right)}{\pi ^{1/2} \Gamma\left(M b\right)}\right]^{\frac{1}{2}}\text{sech}(2Hw)^{M b}.
\end{equation}

By substituting the expression for the warp factor (\ref{scalewarpfactors}) and the expression for $\rho_n(w)$ in (\ref{epsilon_n}) we obtain
\begin{eqnarray}
e_n &=& e\frac{2^{1/4}\sqrt{H}\, \Gamma\left(\frac{1}{4}\right) \Gamma\left(\frac{1}{2}+M b\right)}{\pi^{3/4} \Gamma\left(M b\right)}
{\int} dw\,\text{sech}(2H w)^{2M b-\frac{1}{4}}\left[\sum_{\pm} C_{\pm}(\sigma)
 P_{1/4}^{\pm i\sigma}\left(\text{tanh}(2Hw)\right)\right]\nonumber\\
 &=&e \frac{ \Gamma\left(\frac{1}{4}\right)\Gamma\left(M b-\frac{1}{8}\right)\Gamma\left(\frac{1}{2}+M b\right)}{2^{3/4}\pi ^{1/4}\sqrt{H}\, \Gamma\left(M b\right) \Gamma\left(\frac{3}{8}+M b\right)}
\left[\sum_{\pm} C_{\pm}(\sigma) P_{1/4}^{\pm i\sigma}\left(0\right)\right],
\label{enIV}
\end{eqnarray}
where now we have applied the following definition for the normalized Dirac delta function
\begin{eqnarray}
\delta (w) = \lim_{H\to\infty} ~\frac{2H\Gamma\left(\frac{3}{8}+M b\right)}{\pi ^{1/2} \Gamma\left(M b-\frac{1}{8}\right)}
\text{sech}(2Hw)^{2M b - \frac{1}{4}}, \qquad M b>\frac{1}{8} \label{delta3}
\end{eqnarray}
in the thin brane limit when $H\to\infty$.
The above result leads us to the following form of the Coulomb potential 
\begin{eqnarray}
 &V(r)&\!=\! \frac{e^2}{4{\pi}r}\!\!\left[\!1\! + \!\frac{1}{2^{\frac{3}{2}}\sqrt{\pi } H}\!\!\left(\!\!\frac{\Gamma\!\left(\frac{1}{4}\right)\!\Gamma\!\left(M b\!-\!\frac{1}{8}\right)\!
 \Gamma\!\left(\frac{1}{2}\!+\!M b\right)}{\!\Gamma\!\left(M b\right) \!\Gamma\!\left(M b\!+\!\frac{3}{8}\right)}\!\!\right)^2 \!\!\!\!\int_{m_0}^{\infty}
\!\!\!\!\!\!dm\ \!\text{e}^{-m r} \left|\sum_{\pm} C_{\pm}(\sigma) P_{1/4}^{\pm
i\sigma}\left(0\right)\right|^2\right]
\noindent\nonumber\\
      \!\!&=&\!\!\!\!\! \frac{e^2}{4{\pi}r}\!\!\left[\!1\! + \!\frac{1}{\sqrt{2\pi\, } H} \!\!\left(\!\!\frac{\Gamma\!\left(\frac{1}{4}\right)\!\Gamma\!\left(M b\!-\!\frac{1}{8}\right)\!\Gamma\!\left(\!\frac{1}{2}\!+\!Mb\!\right)}
      {\!\Gamma\!\left(M b\right) \!\Gamma\left(M b\!+\!\frac{3}{8}\right)}\!\!\right)^2 \!\!\!\!\int_{m_0}^{\infty}
\!\!\!\!\!\!dm\ \!\text{e}^{-m r}\!
\left|\frac{\Gamma\left(\!1\!+\!i\sigma\right)}{\!\Gamma\!\left(\frac{3}{8}\!-\!\frac{i\sigma}{2}\right)\!\Gamma\!\left(\frac{9}{8}\!-\!\frac{i\sigma}{2}\right)}\right|^2\right]\!.
\noindent\label{CCl_Gammas3}
\end{eqnarray}
After replacing the integration constants $\left|C_{\pm}(\sigma)\right|$ and using the formula (\ref{Legendre_0}) in the expression for the Coulomb potential (\ref{CCl_Gammas3}), it can be written in the form of (\ref{VCCl2}), where the correction $\Delta V$ now reads
\begin{eqnarray}
 \Delta V =\frac{1}{\sqrt{2\pi}}\left(\frac{\Gamma\left(\frac{1}{4}\right)\Gamma\left(M b-\frac{1}{8}\right)\Gamma\left(\frac{1}{2}+M b\right)}
 {\Gamma\left(M b\right) \Gamma\left(M b\!+\!\frac{3}{8}\right)\Gamma \left(\frac{3}{8}\right) \Gamma\left(\frac{9}{8}\right)}\right)^2 
\frac{e^{-Hr/2}}{Hr}\left(1+{\cal O}\left(\frac{1}{Hr}\right)\right). \label{CCptl3}
\end{eqnarray}

Thus, for all the above considered cases, the corrections to Coulomb's law are exponentially suppressed in the thin brane limit $H\to\infty$. This result is due to the existence of a mass gap in the spectrum of KK gauge field excitations reported in \cite{AHAgauge}.

\subsection{Corrections to Coulomb's law in a thick brane scenario, case IV }

At this point the corrections made for the different cases discussed above are valid only in the limit of thin branes, in which we assumed that the first massive mode of the gauge bosons $m=\frac{H}{2}$ predicted in \cite{AHAgauge} is very large.
However, in some cases we can also analyze the corrections to Coulomb's law from another more realistic point of view, i.e. within another valid approximation for thick brane scenarios. 

Let us start by performing the integral (\ref{enIV}) for $e_{n}$ without the thin brane limit assumption
 \begin{eqnarray}
e_n &=& e\frac{2^{1/4}\sqrt{H}\, \Gamma\left(\frac{1}{4}\right) \Gamma\left(\frac{1}{2}+M b\right)}{\pi^{3/4}\, \Gamma\left(M b\right)}
{\int} dw\,\text{sech}(2H w)^{2M b-\frac{1}{4}} \nonumber\\
&\times& \left[\sum_{\pm} C_{\pm}(\sigma) P_{1/4}^{\pm i\sigma}\left(\text{tanh}(2Hw)\right)\right]\,.
\end{eqnarray}
In order to facilitate this integration it is convenient to introduce the following variable $w=\frac{\text{arctanh}(x)}{2H}$, which leads to  
\begin{eqnarray}\label{int5d}
e_n &=& e\frac{\Gamma\left(\frac{1}{4}\right) \Gamma\left(\frac{1}{2}+M b\right)}{\sqrt{H}\,(2\pi)^{3/4}\, \Gamma\left(M b\right)}
{\int} dx\,\left(1-x^2\right)^{M b-\frac{9}{8}}
\left[\sum_{\pm} C_{\pm}(\sigma)
 P_{1/4}^{\pm i\sigma}\left(x\right)\right].
\end{eqnarray}
When we integrate over the entire fifth dimension using formula $\text{ET II 316(6})$ of the Gradshteyn and Ryzhik handbook \cite{Gradshteyn}, 
the expression for (\ref{int5d}) results in
  \begin{eqnarray}\label{int52d}
 &e_n& = \!e\frac{\Gamma\left(\frac{1}{4}\right) \Gamma\left(\frac{1}{2}+M b\right)}{\sqrt{H}\,(2\pi)^{3/4}\, \Gamma\left(M b\right)} 
{\int^{1}_{-1}} dx\,\left(1-x^2\right)^{M b-\frac{9}{8}}\left[\sum_{\pm} C_{\pm}(\sigma) P_{1/4}^{\pm i\sigma}\left(x\right)\right] = \nonumber\\
 & & \!\!
 e\frac{\Gamma\left(\frac{1}{4}\right) \Gamma\left(\frac{1}{2}+M b\right)}{\sqrt{H}\,(2\pi)^{3/4}\, \Gamma\left(M b\right)}
 \!\left(\! \sum_{\pm}\! C_{\pm}(\sigma)
 \frac{\pi\,2^{\pm i\sigma}\Gamma(2Mb\!-\!\frac{1}{4}\!\pm\! i\frac{\sigma}{2})\Gamma(2Mb\!-\!\frac{1}{4}\!\mp\! i\frac{\sigma}{2})}{\Gamma(2Mb\!-\!\frac{3}{8})\Gamma(2Mb\!+\!\frac{3}{8})\Gamma(\frac{3}{8}\!\mp\! i\frac{\sigma}{2})\Gamma(\frac{9}{8}\!\mp\! i\frac{\sigma}{2})}\!\right), 
\end{eqnarray}
where $M b>\frac{1}{8}$. We then need to square the couplings $e_{n}$ and integrate this expression over all continuous KK massive modes along the lines of (\ref{Potcomplet})
\begin{equation}
 \int^{\infty}_{m_{0}} dm |e_{n}|^2 e^{-mr}.
\end{equation}
Before making this calculation, we will perform the following change of variable $m=\frac{H}{2} \sqrt{1+16 \sigma ^2}$, then the above integral reads
\begin{eqnarray}
& &e^{2}\frac{2^{3/2}\,\Gamma\left(\frac{1}{4}\right)^{2} \Gamma\left(\frac{1}{2}+M b\right)^{2}}{\pi^{3/2}\, \Gamma\left(M b\right)^{2}}
\int_{0}^{\infty} d\sigma\ \frac{e^{-\frac{1}{2}Hr \sqrt{1+16 \sigma ^2}} }{\sqrt{\frac{1}{\sigma ^2}+16 }}\times \nonumber\\
& &\left|\sum_{\pm} C_{\pm}(\sigma)
 \frac{\pi\,2^{\pm i\sigma}\Gamma(2Mb-\frac{1}{4}\pm i\frac{\sigma}{2})\Gamma(2Mb-\frac{1}{4}\mp i\frac{\sigma}{2})}{\Gamma(2Mb-\frac{3}{8})\Gamma(2Mb+\frac{3}{8})\Gamma(\frac{3}{8}\mp i\frac{\sigma}{2})\Gamma(\frac{9}{8}\mp i\frac{\sigma}{2})}  \right|^2\ .
   \label{int_m}
\end{eqnarray}
It seems impossible to do this integral analytically, however, as we have previously assumed when computing (\ref{CCptl2}), we shall consider that the contribution to the mass integral (\ref{int_m}) is dominated by the first KK continuous excitation modes. Therefore we can expand the prefactor of the exponential around $\sigma=0$, using $C_\pm(\sigma)=\frac{\left|\Gamma(1\mp i\sigma)\right|}{\sqrt{2\pi}}$, and we obtain the following approximate result for the above integral
\begin{eqnarray}\nonumber
& &\!\!\!e^{2}\frac{2^{3/2}\, \Gamma\left(\frac{1}{4}\right)^{2} \Gamma\left(\frac{1}{2}+M b\right)^{2}}{\pi^{3/2}\, \Gamma\left(M b\right)^{2}}\!\!\int_{0}^{\infty}
\!\!\!\!d\sigma\ \!\!\frac{e^{-\frac{1}{2}Hr \sqrt{1+16 \sigma ^2}} }{\sqrt{\frac{1}{\sigma ^2}+16 }} \times \\
& &\left|\sum_{\pm} C_{\pm}(\sigma)
 \frac{\pi\,2^{\pm i\sigma}\,\Gamma\!(2Mb\!-\!\frac{1}{4}\!\pm\! i\frac{\sigma}{2})\Gamma\!(2Mb-\frac{1}{4}\!\mp\! i\frac{\sigma}{2})}{\Gamma\!(2Mb\!-\!\frac{3}{8})\Gamma\!(2Mb\!+\!\frac{3}{8})\Gamma\!(\frac{3}{8}\!\mp\! i\frac{\sigma}{2})\Gamma\!(\frac{9}{8}\!\mp\! i\frac{\sigma}{2})}  \right|^2\ \!\!\!\!\sim\\\nonumber 
& &\!e^{2}\left(\frac{2^6\, \Gamma\!\left(\frac{1}{4}\right)^{2} \Gamma\!\left(\frac{1}{2}\!+\!M b\right)^{2} \Gamma \!\left(2Mb\!-\!\frac{1}{4}\right)^4}{(2\pi)^{\frac{1}{2}}\Gamma \!\left(\frac{3}{8}\right)^2 \Gamma \!\left(\frac{1}{8}\right)^2 \Gamma\!\left(M b\right)^{2} \Gamma \!\left(2Mb-\frac{3}{8}\right)^2 \Gamma \!\left(2Mb+\frac{3}{8}\right)^2} \!\right)\frac{\text{e}^{-\frac{1}{2} H r}}{Hr}
\!\left(\!1+\!{\cal O}\left(\frac{1}{Hr\!}\right)\!\right),
\end{eqnarray}
finally the  expression for the corrected Coulomb's potential reads
\begin{eqnarray}\label{c4a}
 V = \frac{e^{2}}{4\pi r}\left[1+ \Delta V \right],
\end{eqnarray}
where the correction is given by
\begin{eqnarray}\label{c4}
\Delta V = \gamma\left(Mb\right)\frac{e^{-\frac{1}{2} H r}}{ H r}
\left(1+{\cal O}\left(\frac{1}{Hr}\right)\right),
\end{eqnarray}
and the constant function $\gamma(Mb)$ explicitly depends on the 5D parameters $M$ and $b$ and possesses the form 
\begin{equation}\nonumber
\gamma\left(Mb\right)=\left(\frac{8~\Gamma\left(\frac{1}{4}\right) \Gamma\left(\frac{1}{2}+Mb\right) \Gamma \left(2Mb-\frac{1}{4}\right)^2}
{(2\pi)^{\frac{1}{4}}\Gamma \left(\frac{3}{8}\right) \Gamma \left(\frac{1}{8}\right) \Gamma\left(Mb\right) \Gamma \left(2Mb-\frac{3}{8}\right) \Gamma \left(2Mb+\frac{3}{8}\right)} \right)^{2}.
\end{equation}
This result is particularly relevant when directly compared to the one obtained in the thin brane limit (\ref{CCptl3}), because the corrections to Coulomb's law hold their shape for large and small mass scales dictated by the Hubble parameter $H$ since the first excited state possesses $m=H/2$.
Furthermore, we can set the value of the constant $\gamma(Mb)\sim H$ implying that the product $Mb\sim 5$. Thus, by making this assumption we ensure that the 5D parameter $b$ (which depends on $\Lambda_5$ and $\kappa_5$) and the coupling constant $M$ have the same order of magnitude, simplifying indeed the potential as follows
\begin{eqnarray}\label{c4b}
 V = \frac{e^{2}}{4\pi r}\left(1+ \frac{e^{-\frac{1}{2} H r}}{ r} \right).
\end{eqnarray}
It is worth mentioning that the corrections to this potential are still far beyond the upper experimental bound observed on the photon mass, making the present braneworld model phenomenologically viable. In fact, by confronting our results to actual experimental observations, we can consider a realistic scenario by setting the Hubble parameter to its present value $H=H_0=1\times 10^{-33}\, \text{eV}$, leading to an estimation of the order $m_\gamma\sim 5\times 10^{-34}\, \text{eV}$, while the experimental estimation reported by Ryutov is of the order $m_\gamma\sim 5\times 10^{-18}\, \text{eV}$ according to \cite{terrestial}-\cite{GN}.



\section{Conclusions and discussion}

With this work we contribute to the program of localization of various matter fields in the tachyonic de Sitter thick braneworld constructed in \cite{HRMGR}. The primary goal was to localize fermion fields in the 3--brane since gravity, as well as scalar and gauge vector fields were already localized in this expanding braneworld scenario. A second goal consisted in computing the corrections to Coulomb's law coming from the extra dimensional nature of the KK massive gauge vector excitations. These aims were successfully achieved for three different discussed cases.

This work was carried out by considering four different functions $F(T)$, which establish the concrete 5D Yukawa interaction between fermions and the tachyon condensate scalar field.

In the first case we set $F(T)$ proportional to the field $T$. However, for this configuration neither left-- nor right--chiral fermions are localized in the 3--brane (although the left Schr\"odinger potential is of volcano type) since the left--chiral zero mode asymptotically tends to a positive constant and is therefore delocalized from it. 

For case II we have field configurations endowed with left and right Schr\"odinger potentials with infinitely high walls which have discrete mass spectra for the KK modes. The left--chiral fermionic massless zero mode, as well as an infinite number of discrete massive bound states are localized on the brane, whereas the right--chiral zero mode is delocalized from it.

For the case III we found a Schr\"odinger potential of volcano type for left--chiral fermions, where only the ground state corresponding to the KK massless zero mode is localized and glued to the continuous KK massive spectrum. On the other hand, the corresponding right Schr\"odinger potential does not localize any right--chiral fermions on the 3--brane. 

In the case IV we got modified P\"oschl--Teller potentials with mass gaps which allows us to localize both left-- and right--chiral fermions on the brane and to get discrete KK mass spectra where the left--chiral massless zero mode is separated from the continuous massive spectrum of KK excitations; for this scenario the right--chiral KK massless zero mode is non--localized on the 3--brane.

As mentioned above, after localizing the fermion fields, our second objective was to make use of the results presented here and in \cite{AHAgauge}, which show that it is possible to localize gauge fields in our braneworld model, in order to study the interaction between photons and fermions localized on the brane. We further performed the computation of the corrections to the Coulomb's law coming from the massive gauge vector modes by considering the aforementioned cases II, III and IV. The computed corrections to the Coulomb's potential exponentially decay due to the presence of a mass gap in the spectrum of the gauge vector fields. Thus, these corrections decay much faster than $1/r$ due to the exponential function that quickly removes all small contributions from the KK massive gauge vector modes in the limit of thin branes. 


Moreover, for case IV, it is possible to obtain a novel result that displays the corrections to Coulomb's law in a tachyonic de Sitter braneworld scenario of arbitrary thickness, allowing us to get an idea of what would be the effects of the electromagnetic interaction between localized fermions, due to non--localized massive gauge bosons. When confronting the estimated correction to the Coulomb law with the experimental upper bound on the photon mass we find the our prediction is far away from being detected in the near future.

\acknowledgments
AHA is grateful to ICF, UNAM and UAM-I for hospitality, as well as to ``Programa de Apoyo a Proyectos de Investigaci\'on e
Innovaci\'on Tecnol\'ogica'' (PAPIIT) UNAM, IN103413-3, Teor{\'i}as de Kaluza-Klein, inflaci\'on y perturbaciones gravitacionales. 
RRML acknowledges a postdoctoral grant from CONACyT at ICF-UNAM. All authors thank SNI for support.



\end{document}